\documentclass[showpacs,twocolumn,prl]{revtex4-1}

\usepackage{amsmath}
\usepackage{amsfonts}
\usepackage{amssymb}
\usepackage{graphicx}
\usepackage{subfigure}

\begin{document}

\title{Anomalous diffusion in run and tumble motion}

\author{Felix Thiel}\affiliation{Institute of Physics,
  Humboldt University Berlin, Newtonstr. 15, 12489 Berlin, Germany}

\author{Lutz Schimansky-Geier}\affiliation{Institute of Physics, Humboldt
  University Berlin, Newtonstr. 15, 12489 Berlin, Germany}

\author{Igor M. Sokolov}\affiliation{Institute of Physics, Humboldt
  University Berlin, Newtonstr. 15, 12489 Berlin, Germany}

\date{\today}

\begin{abstract}
  A random walk scheme, consisting of alternating phases of regular Brownian motion and
  L\'evy-walks, is proposed as a model for run-and-tumble
  bacterial motion. Within the continuous-time random walk approach we obtain the long-time and
  short-time behavior of the mean squared displacement of the walker
  as depending on the properties of dwelling time distribution in each phase.  
  Depending on these distributions, normal diffusion, superdiffusion and ballistic spreading
  may arise.
\end{abstract}
\maketitle


We propose a qualitative model for the movement of motile
bacteria like E. Coli.  These organisms have several flagella, by
which they are propelled.  The motion of E. Coli can be subdivided in two
different phases: tumble- and run-mode.  In the tumbling-mode the flagella 
rotate in different directions, causing an erratic movement
and a fast change of orientation.  In the run-mode all flagella
rotate in the same direction and entangle, resulting in a straight
forward movement with a constant velocity.  The bacteria use the
tumbling phase for chemotaxis, swimming in the direction of nutrients
or away from poisonous areas.  The whole process is described for
instance in \cite{Berg2004} and its numerical and theoretical
description is a subject of recent discussion, see e.g.
\cite{Othmer1988, Visser2003, NicolauJr2009}.

Random walks are the model of choice in description of motion of
living organisms \cite{Berg2004, Othmer1988, NicolauJr2009,
  Meysman2008}.  Run and tumble motion is often described either by a
L\'evy-walk or a continuous time random walk (CTRW), also called
``velocity'' and ``jumping process'' \cite{Othmer1988}.  CTRWs
describe a random process consisting in instantaneous jumps separated
by periods of rest (waiting times) following a given probability
distribution as first discussed in \cite{Montroll1965}.  A L\'evy-walk
is a motion at a constant speed where the direction of velocity is
changed after random epochs \cite{Shlesinger1987}.  It is essentially
a CTRW in which the waiting period corresponds to a motion with
constant velocity and jumps lead to a change of the direction of
motion. Although L\'evy-walks are commonly used to describe the (random) 
motion of bacteria, the origin of this kind of motion is a topic of 
recent studies (see e.g. \cite{Matthaeus2009}).

In this paper we combine both processes and model the run-and-tumble motion as an
alternating random walk consisting of phases of simple
diffusion (tumble phase) and of L\'evy walk (run phase). A sketch
of the overall process is shown in Figure 1. The speed of motion in
the run phase will be denoted by $c$ and the diffusion coefficient in
the tumbling phase by $D$. The dwelling time $t$ in each phase is
given by waiting time probability density functions (PDFs) $\psi_1(t)$
and $\psi_2(t)$ for the tumbling and run phases, respectively. All
other properties in the run and tumbling phases will be denoted by
subscripts $1$ and $2$ in what follows. The whole motion takes place
in the space of $d$ dimensions, with $d=2$ or $3$.  For the sake of
simplicity, we assume that the observation starts at a beginning of a
run or of a tumble phase (in order to avoid the discussion of aging effects).  
Although the alternating setup is not a new process in
the theory of random walks, and the general discussion follows the
standard lines (see \cite{Klafter2011}), the model discussed below is
new and far from being trivial. 

\begin{figure}
  \includegraphics[width=0.45\textwidth]{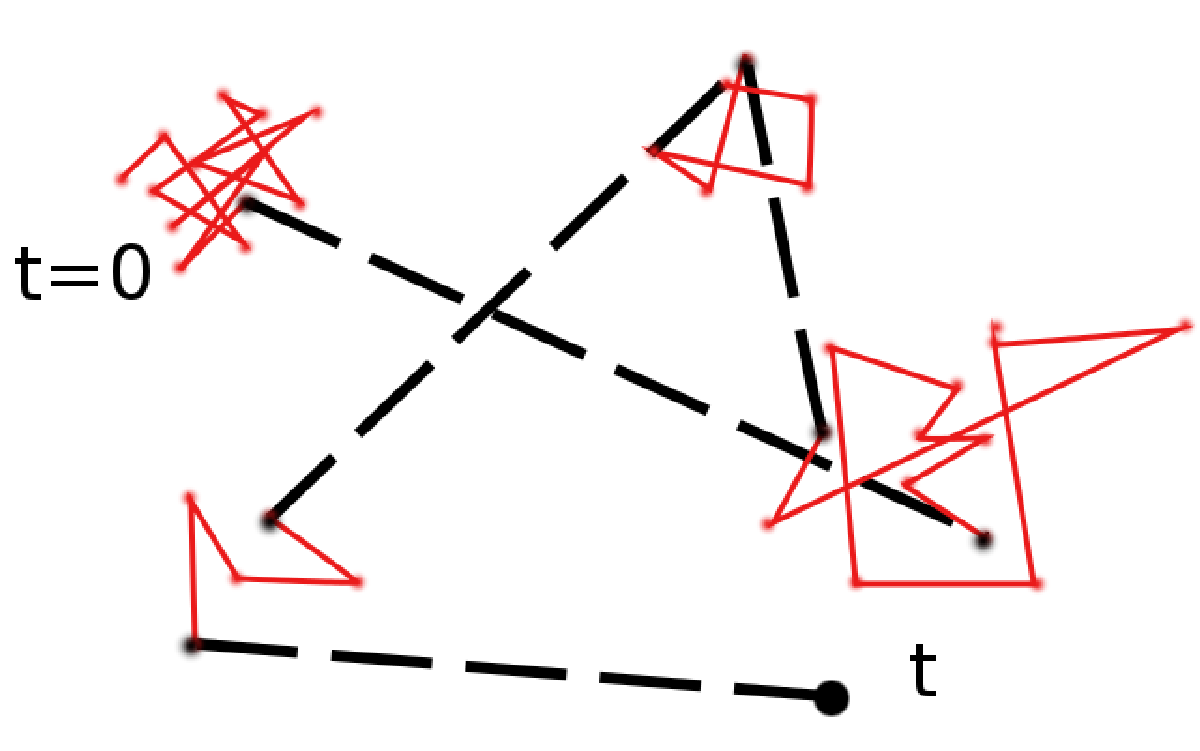}
  \begin{caption}{Sketch of the trajectory. Tumbling motion is
      displayed with solid lines and alternates with L\'evy walk
      stretches (dashed lines).  After a certain time the process is
      terminated, here in the run phase.}
	\end{caption}
	\label{fig:1}
\end{figure}

The displacements $\mathbf{x}$ in a tumbling phase are given by the
transition probability density of Brownian motion
\begin{equation}
	\lambda_{1}(\mathbf{x}|t) = \frac{1}{ \left( 4 \pi D t \right)^{\frac{d}{2}}} \exp \left( - \frac{ \mathbf{x}^2 }{4 D t } \right)\,.
\end{equation}
It depends on the time $t$ the walker has spent in this
phase. Respectively, if $t$ is the elapsed time in the run phase, the
transition probability density for a walk over distance $x$ reads:
\begin{equation}
  \lambda_{2}(\mathbf{x}|t) = \frac{1}{S_{d-1}( c t)^{d-1}}\delta \left(|\mathbf{x}| - c t \right)\,.
\end{equation}
Here $S_{d-1}$ is the (hyper)surface of a $d$-dimensional unit
ball which is given by: $S_{d-1} = (2 \pi^{d/2})/\Gamma(d/2)$, with
$\Gamma(x)$ being the Gamma function.  The corresponding probability
densities are isotropic as long as no chemotactic effects are
included.

We discuss the properties of the propagator $G(\mathbf{x}_1, t_1 |
\mathbf{x}_0, t_0)$ of the complete process starting at $t=0$ and
stopped (observed) at time $t$, and concentrate on the mean square displacement (MSD)
\begin{equation}
  \langle\mathbf{x}^2(t)\rangle = \int\mathbf{x}^2 G(\mathbf{x}, t |0,0) d \mathbf{x}\,.
\end{equation}
Typically, the MSD follows a
power law pattern, $\langle\mathbf{x}^2(t)\rangle \propto t^{\gamma}$;
the value $\gamma = 1$ corresponds to normal diffusion and cases with
$\gamma > 1$ to superdiffusion.  The special limit $\gamma = 2$ is
referred to as ballistic spreading. Subdiffusion, with $0<\gamma < 1$,
is impossible in our setting. Since our situation is homogeneous both in space and in time, it is
easier to use the Fourier representation for the coordinate and the Laplace representation
for time in which $G$ is given by an algebraic expression.

Typically, the process is stopped not at a time it changes the
phase. Hence, we need two additional transition probability densities,
which are
\begin{equation}
  \Lambda_i(\mathbf{x},t) = \lambda_i(\mathbf{x}|t) \int_t^\infty \psi_i(t') dt'\,.
\label{Eq:4}
\end{equation}
They describe the displacement in a prematurely finished phase
($i=1,2$).  The integral in this expression gives the probability that
the dwelling time in phase $i$ exceeds $t$.  Its Laplace transform is
$(1 - \psi_i (s))/s$.  Thus, the $\Lambda_i$ are the complete
analogons to the regular CTRWs waiting time PDFs after the last
jump.
Eq. (\ref{Eq:4})
takes into account the fact that the walker continuously moves during the
waiting period.

Let us assume that walks start with the beginning of a tumble phase.  
To find the propagator, we note that the corresponding
process may have completed some number $m \ge 0$ of full tumble and
run cycles before the observation at time $t$ took place, and that the
time is stopped either during the tumble or during the following run
phase of the last, incomplete, cycle. A propagator corresponding to a full cycle
is simply a convolution of $\sigma_1(\mathbf{x},t) = \lambda_1(\mathbf{x}|t) \psi_1(t)$ and 
$\sigma_2(\mathbf{x},t) = \lambda_2(\mathbf{x}|t) \psi_2(t)$.
In a Fourier-Laplace domain $\sigma_i(\mathbf{x},t)$ transforms to 
$\sigma_i(\mathbf{k},s)$. The Fourier-Laplace representation of the
propagator for $m$ completed cycles is given by $[\sigma_1(\mathbf{k},s)\sigma_2(\mathbf{k},s)]^m$.  
The remaining last cycle can be stopped either in the tumble phase or in the run  phase.
In the first case, the joint PDF of displacement and time in this last phase
will be given by $\Lambda_1(\mathbf{k},s)$, or, in the Fourier-Laplace
representation, by $\Lambda_1(\mathbf{k},s)$. If the incomplete cycle
finishes in the run phase, this PDF is given by the convolution of
$\sigma_1(\mathbf{x},t)$ and $\Lambda_2(\mathbf{x},t)$, corresponding to $\sigma_1(\mathbf{k},s)
\Lambda_2(\mathbf{k},s)$ in the Fourier-Laplace representation.
Summing up all the possibilities,  we get 
\begin{align}
G_1&(\mathbf{k},s) =\\
& = \sum_{m=0}^\infty [\sigma_1(\mathbf{k},s)\sigma_2(\mathbf{k},s)]^m [\Lambda_1(\mathbf{k},s)+ \sigma_1(\mathbf{k},s) \Lambda_2(\mathbf{k},s)], \nonumber
\end{align}
yielding for walkers starting with a tumbling phase  
\begin{align}
G_1(\mathbf{k},s) = \frac{ \Lambda_{1}(\mathbf{k},s) +\sigma_{1}(\mathbf{k},s)  \Lambda_{2}(\mathbf{k},s)}{1 - \sigma_{1}(\mathbf{k},s)   \sigma_{2}(\mathbf{k},s) }.\nonumber
\end{align}
Since the walker could have started in both phases, we average
over the two possibilities. The averaged propagator reads
\begin{equation}
\label{eq:1}
G (\mathbf{k},s) = \frac{ \Lambda_{1} + \Lambda_2 + \sigma_{1}  \Lambda_{2} + \sigma_2 \Lambda_1 }{2 (1 - \sigma_{1}  \sigma_{2}) }\,,
\end{equation}
where we have omitted the arguments of the corresponding functions. Due
to its symmetry, this expression is easier to handle and will be used
for the further analysis.

The MSD in the Laplace domain is given by
\begin{equation}
  \langle\mathbf{x}^2 (s)\rangle = - \Delta_{\mathbf{k}} \left. G(\mathbf{k},s) \right|_{\mathbf{k} = 0}
\end{equation}
where $\Delta_{\mathbf{k}}$ denotes the Laplacian in the Fourier
space. To relate this MSD to the PDFs of the waiting times and to the
transition probabilities, we perform explicitly the Fourier-Laplace
transforms of $\sigma_{1,2}$ and $\Lambda_{1,2}$. It gives in case of
tumbling:
\begin{align}
  \nonumber \sigma_{1}(\mathbf{k},s) & = \int_0^\infty dt e^{-st} \int d \mathbf{x} e^{{\rm i}\mathbf{k}\mathbf{x}}  \lambda_1(\mathbf{x}| t) \psi_1(t) \\
  \nonumber & = \int_0^\infty dt e^{-st - D \mathbf{k} ^2} \psi_{1}
  (t) = \psi_{1} (s + D \mathbf{k} ^2)
\end{align}
where $\psi_{1}(...)$ is the Laplace representation of $\psi_1(t)$. The
same procedure applied to $\Lambda_1(\mathbf{x},t)$ gives
\begin{equation}
	\Lambda_{1} (\mathbf{k},s) = \frac{1 - \psi_{1} (s + D \mathbf{k}^2 ) }{s + D \mathbf{k}^2 }.
\end{equation}
Let us now turn to the L\'evy walk phase. It reads
\begin{align}
  & \sigma_{2}(\mathbf{k},s)  = \\
  &=\,\int_{\Omega_{d-1}} d \mathbf{e} \int_0^\infty dr r^{d-1}
  \int_0^\infty dt \frac{\delta(r - c t) \psi_2 (t)}{S_{d-1} (c
    t)^{d-1}} e^{ - s t + i r \mathbf{k} \mathbf{e}} \,. \nonumber
\end{align}
Performing integration in $r$ we obtain
\begin{align}
  \sigma_{2}(\mathbf{k},s) = \frac{1}{S_{d-1}}\int_{\Omega_{d-1}} d\mathbf{e} \psi_{2} (s - i c \mathbf{k}\mathbf{e})
\end{align}
Here $\Omega_{d-1}$ denotes the hypersurface of a $d$-dimensional unit
ball, and $\mathbf{e}$ is an unit vector defining the point on this
surface. Integration in $\mathbf{e}$ corresponds in $d=2,3$ to the
angle-integration or over the solid angle. Analogously,
\begin{equation}
	\Lambda_{2} (\mathbf{k},s) = \frac{1}{S_{d-1}} \int_{\Omega_{d-1}} d \mathbf{e} \frac{1 - \psi_{2} (s - i c \mathbf{k} \mathbf{e})}{s - i c \mathbf{k} \mathbf{e}}\,.
\end{equation}

Evaluating the Laplacian of the propagator one obtains a lengthy
expression involving $\sigma_{1,2}$, $\Lambda_{1,2}$ and their first
and second partial derivatives taken at $\mathbf{k}=0$. All items are
expressed through $\psi_{1,2}$ and their first and second derivatives
with respect to their argument. For the running phase, they have
additionally to be integrated over the surface of the unit ball. All
first derivatives of $\sigma_{1,2}$ and $\Lambda_{1,2}$ vanish due to
spacial symmetry.  Terms containing a second derivative of
$\psi_{1,2}$ enter expressions for $\sigma_{1,2}$ and for
$\Lambda_{1,2}$ with opposite signs and cancel. The compact final
result consists of two parts corresponding to the running and the tumbling
phase. It reads:
\begin{align}
\label{eq:2}
	\langle \mathbf{x}^2 (s) \rangle &=  \frac{ d D (1 + \psi_2)}{s \left( 1 - \psi_{1} \psi_{2} \right)} \frac{1 - \psi_1}{s}\\
	\nonumber	& + \frac{c^2 (1 + \psi_{1})}{s \left( 1 - \psi_{1} \psi_{2} \right)} \left(\frac{\psi_{2}'}{s} + \frac{1 - \psi_{2}}{s^2} \right), 
\end{align}
where the derivative $\psi'$ is taken with respect to the
Laplace-variable $s$. 

\textbf{Exponential Waiting Time PDFs.} We now proceed by making
specific assumptions about the waiting time PDFs $\psi_i$. We first
assume that all waiting time PDFs take exponential forms so that
$\psi_{i} (s) = 1/(s \tau_{i} + 1)$ with $\tau_i$ being mean dwelling
times in the corresponding phases.  By plugging these expressions into
(\ref{eq:2}), we can get an expression for the second moment in the
spectral domain
\begin{align}
 & \langle \mathbf{x}^2 (s)\rangle = \\
& = \frac{ d D \tau_1 (s \tau_2 + 2)}{s^2 (s \tau_1 \tau_2 + \tau_1 + \tau_2)} +  \frac{c^2 \tau_{2}^{2} (s \tau_1 + 2)}{s^2 (s \tau_2 + 1) (s \tau_1 \tau_2 + \tau_1 + \tau_2)}\,, \nonumber
\end{align} 
and from that, the asymptotic scaling in the time domain
\begin{align}
  \langle \mathbf{x}^2 (t)\rangle	\simeq d D t ~~~~~~~~~~~~~~ &\qquad \mbox{for} \qquad t \rightarrow 0\,, \\
  \langle \mathbf{x}^2 (t)\rangle \simeq 2 \frac{d D \tau_1 + c^2
    \tau_{2}^{2}}{\tau_1 + \tau_2} t &\qquad \mbox{for} \qquad t
  \rightarrow \infty\,. \label{LoTi}
\end{align}
In order to obtain this result, we used Tauberian theorems, which
state that the small $s$ limit corresponds to a large $t$ limit in
original domain and vice versa.  In both limits $t \rightarrow 0$ and
$t \rightarrow \infty$, one observes normal diffusion, albeit with
different diffusion coefficients. 

\textbf{Power Law PDFs.}  We will now consider waiting time
distributions asymptotically following power laws for $t
\rightarrow \infty$:
$\psi_1(t) \propto t^{- \left( 1 + \alpha \right) }$ and $\psi_2(t)
\propto t^{- \left( 1 + \beta \right) }$. Interesting new features
arise if $\alpha \leq 1$ and $\beta \leq 2$. If the exponents are 
equal or larger than unity, the first moments of the waiting times are finite 
but the second ones diverge. If the exponents are less than unity, i.e.
$\alpha, \beta < 1$, even the means do not exist.
As we proceed to show, the diffusive behavior is governed by the first 
non-analytic term of $\psi_{1}(s)$, resp. $\psi_{2}(s)$.

Let us first discuss the case when the first moments of both dwelling time distributions diverge. 
The corresponding expansions of the Laplace-transforms in the limit $s
\rightarrow 0$ read
\begin{align}
  \psi_{1}(s) & = 1 - A_1 s^{\alpha} + \dotsb \label{PsiOne}\\
  \psi_{2}(s) & = 1 - A_2 s^{\beta} + \dotsb \label{PsiTwo}
\end{align} 
with $\alpha, \beta < 1$. In this case Eq.(\ref{eq:2}) gives in the leading
order
\begin{equation}
	\left\langle \mathbf{x}^2 (s) \right\rangle \approx \frac{2 A_2 c^2 \left( 1 - \beta \right) s^{\beta - 2} }{A_1 s^{\alpha + 1} + A_2 s^{\beta + 1}}\,.
\end{equation}
Remarkably, this asymptotic expression does not explicitly depend on
the diffusion coefficient $D$ of the tumbling phase. Transformed back
to the time domain, the diffusive scaling depends on the relation between
$\alpha$ and $\beta$:
\begin{equation}
	\langle\mathbf{x}^2 (t) \rangle \simeq c^2 (1 - \beta) \left\{
	\begin{array}{l l}
		t^2 &\mbox{for} \; \beta < \alpha \\
		 \frac{A_2}{A_1 + A_2}t^2 &\mbox{for} \;  \beta = \alpha \\
		 \frac{2}{\Gamma(3 + \alpha - \beta )} \frac{A_2}{A_1} t^{2 - (\beta - \alpha)} &\mbox{for} \; \beta > \alpha
	\end{array}
	\right. \label{LeWa}
\end{equation}
Hence, we obtain superdiffusion in the long-time limit, which is
always ballistic for $\beta \leq \alpha$. It is independent on the
properties of the tumbling phase for  $\beta < \alpha$, and depends on 
a prefactors $A_1$ and $A_2$ if $\alpha=\beta$.
The first two regimes are dominated by L\'evy walks.  The third
superdiffusive one with $\beta > \alpha$ is close to a sequence of
L\'evy walks interrupted by rests. The tumbling periods hardly
contribute to the displacement, and, therefore, a dependence on $D$
is still absent.

As one can see, the asymptotic expressions (\ref{LeWa}) fail for
$\beta \geq 1$.  In this case, subleading orders in Eqs.(\ref{PsiTwo})
must be taken into account. With $\beta > 1$, 
the Laplace transform of the waiting time density in the run phase
reads
\begin{equation}
  \psi_{2} = 1 - A_2 s + B_2 s^\beta + \dotsb \text{.}
\end{equation}
$A_2$ stands for the finite mean waiting time in this phase.
The case $\beta = 1$ leads to logarithmic corrections, which do not
change the qualitative behaviour. We now 
consider the case $1 < \beta \leq 2$ when the first moment of the dwelling time in the
run phase does exist, but the second one diverges except for the limiting value $\beta=2$. 
Again we let $0 < \alpha \leq 1$ for the tumble phase. Repeating the procedure, we get
\begin{equation}
	\langle \mathbf{x}^2 (s) \rangle \approx 2 \frac{d D A_1 s^{\alpha - 1} +B_2 c^2 \beta s^{\beta - 2} }{A_1 s^{\alpha + 1} + A_2 s^2} \label{key} 
\end{equation}
wherein an explicit dependence on $D$ appears. The behavior in the
time domain depends on relation between $\alpha$ and $\beta$:
\begin{equation}
	\langle \mathbf{x}^2 (t) \rangle \simeq \left\{
	\begin{array}{l l}
		\frac{2(\beta-1) c^2 }{\Gamma(3 + \alpha - \beta)} \frac{B_2}{A_1} t^{2  - (\beta - \alpha)} &\mbox{for} \; \beta-1 < \alpha < 1\\
		2 \left(d D + c^2 (\beta-1) \frac{B_2}{A_1} \right) t  &\mbox{for} \; \beta-1 = \alpha < 1\\
		2 d D t  &\mbox{for} \;  \alpha < \beta-1 < 1 
	\end{array}
	\right. \label{MiWa}
\end{equation}
If $\beta-1 < \alpha$, the behavior of the 
walker still corresponds to superdiffusion, and the
$D$-dependence is suppressed. This case again is similar to
L\'evy walks interrupted by rests where the tumbling
does not contribute to the displacement.  In the two following
situations with $\beta-1 \geq \alpha$, the walker exhibits a normal
diffusive behavior.  In the last one with $\beta>1+\alpha$, the MSD is
dominated by the random motion in the tumbling phase. The running
phase has no effect on the value of the diffusion coefficient. For
short times $t \to 0$ the motion is always dominated by normal diffusion.

In the limit $\alpha \to 1$ the PDF of waiting
times in the tumbling phase gets a finite first moment. The
corresponding scalings are obtained from Eqs. (\ref{LeWa}) and
(\ref{MiWa}). If $\beta <1$, i.e. if the waiting times in the run phase
do not possess neither a first nor a second moment, the motion is
ballistic. With growing $\beta>1$ the motion tends to a
superdiffusive one. The special limit is when $\alpha =1$ and $\beta=2$
corresponds to the purely diffusive case. Then $\psi_1$ does possess
the first moment $\tau_1$ corresponding to $A_1$ in the expansion
(\ref{PsiTwo}) and $\psi_2$ does possess the second moment with
value $2 \tau_2^2$ expressed by $2 B_2$ in the Laplace transform
(\ref{PsiTwo}).  The MSD can be obtained from the expression
(\ref{key}) by taking now into account the contribution of the second
term in the denominator. This result coincides in the long time
asymptotics with the one for the exponentially distributed waiting times
Eq.(\ref{LoTi}).

We have seen that by choosing proper waiting time
distributions of an alternating CTRW process as described above
one can describe different propagation regimes in the long-time limit 
ranging from normal diffusion up to ballistic spreading. In the case of power-law
waiting time PDFs normal diffusive
behavior can only be achieved if the distribution for the Brownian part 
has a heavier tail than the one for runs ($\alpha \leq \beta-1 \leq 1$), which 
has to possess a first moment, i.e. the PDF of
the Brownian tumbling phase has to have significantly more mass in the
tail than its L\'evy-counterpart.
Such power-law-like behavior of a waiting time distribution is no rarity 
and occurs in several other (particularly biological) contexts  \cite{Mejias2010}.

We therefore expect a transition from normal to superdiffusion in
cases with $\alpha > \beta-1>0 $ or with $\beta < 1$.  Fig. 2 shows
what diffusive regime may be expected for different values of $\alpha$ and $\beta$.

\begin{figure}[t]
  \includegraphics[width = 0.5\textwidth]{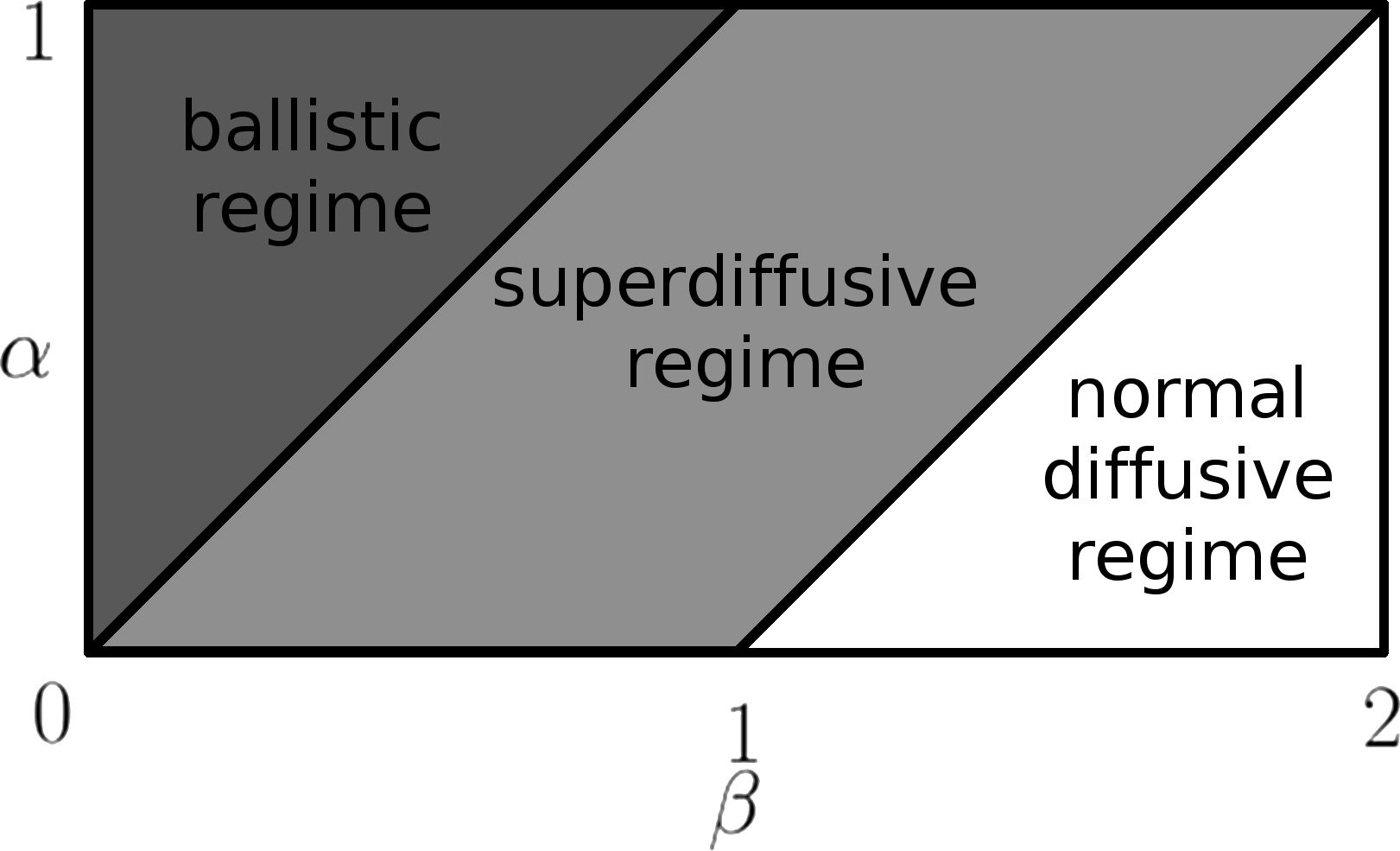}
  \begin{caption}{Phase diagram of the diffusive regimes.  The figure
      shows which diffusive behavior has to be expected in the
      long-time limit. $\alpha$ is the exponent corresponding to the tumbling-phase,
      and $\beta$ is the exponent of the run phase.}
  \end{caption}
  \label{fig:2}
\end{figure}

We note that if the observation does not start at time of a phase
change, the model with the considered power-law distributed waiting
times shows aging effects \cite{Klafter2011}. It may result in changes
of prefactors in the corresponding asymptotic expressions but won't
lead to different asymptotics, so that the classification given in
Fig. 2 still will hold true.

Let us summarize our findings.  We proposed a phenomenological model
for bacteria performing a run-and-tumble motion in the absence of
chemotaxis. We have determined the possible diffusive regimes of such
motion.  Depending on the distribution of times in the two phases of
motion, the model shows transitions from normal diffusive to
superdiffusive and to ballistic behavior.  Our model may prove to be
useful to model situations that exhibit such transitions, such as the
one described in \cite{Metzner2007}.  It also may be suitable to
describe non-biological situations like the motion of nanorods, 
which may also exhibit superdiffusion \cite{Campos2009, Golestanian2009}.

The authors acknowledge financial support by DFG within IRTG 1740 research and
training group project.


\begin{thebibliography}{13}
	\bibitem{Berg2004} H. C. Berg, \textit{E. Coli in Motion} (Springer, Heidelberg, 2004).
	\bibitem{Othmer1988} H. G. Othmer, S.R. Dunbar, and W. Alt, J. Math. Biol. \textbf{26}, 263 (1988).
	\bibitem{Visser2003} A. W. Visser and U.H. Thygesen, J. Plankton Res. \textbf{25}, 1157 (2003).
	\bibitem{NicolauJr2009} D. V. Nicolau Jr., J. P. Armitage, and P. K. Maini, Comput. Biol. Chem. \textbf{33}, 269 (2009).
	\bibitem{Meysman2008} F. J. R. Meysman, V. S. Malyuga, B. P. Boudreau, and J. J. Middelburg, Geochim. Cosmochim. Ac. \textbf{72}, 3460 (2008).
	\bibitem{Montroll1965} E. W. Montroll and G. H. Weiss, J. Math. Phys. \textbf{6}, 167 (1965).
	\bibitem{Shlesinger1987} M. F. Shlesinger, B. J. West, and J. Klafter, Phys. Rev. Lett. \textbf{58}, 1100 (1987).
	\bibitem{Matthaeus2009} F. Matth\"{a}us, M. Jagodi\v{c}, and J. Dobnikar, Biophys. J. \textbf{97}, 946 (2009).
	\bibitem{Klafter2011} J. Klafter and I. M. Sokolov, \textit{First steps in random walks}, 1st ed. (Oxford UP, 2011).
	\bibitem{Mejias2010} J. Mejias, H. J. Kappen, and J. J. Torres, PLoS ONE \textbf{5}, e13651 (2010).
	\bibitem{Metzner2007} C. Metzner, C. Raupach, D. Paranhos Zitterbart, and B. Fabry, Phys. Rev. E \textbf{76}, 021925 (2007).
	\bibitem{Campos2009} D. Campos and V. M\'{e}ndez, J. Chem. Phys. \textbf{130}, 134711 (2009).
	\bibitem{Golestanian2009} R. Golestanian, Phys. Rev. Lett. \textbf{102}, 188305 (2009).
\end{thebibliography}

\end{document}